# Modeling correlated noise is necessary to decode uncertainty

R.S. van Bergen[1,*] & J.F.M. Jehee[1,*]

[1]Donders Institute for Brain, Cognition and Behavior, Radboud University, Nijmegen, the Netherlands. *Correspondence should be addressed to r.vanbergen@donders.ru.nl or janneke.jehee@donders.ru.nl

**Abstract:** Brain decoding algorithms form an important part of the arsenal of analysis tools available to neuroscientists, allowing for a more detailed study of the kind of information represented in patterns of cortical activity. While most current decoding algorithms focus on estimating a single, most likely stimulus from the pattern of noisy fMRI responses, the presence of noise causes this estimate to be uncertain. This uncertainty in stimulus estimates is a potentially highly relevant aspect of cortical stimulus processing, and features prominently in Bayesian or probabilistic models of neural coding. Here, we focus on sensory uncertainty and how best to extract this information with fMRI. We first demonstrate in simulations that decoding algorithms that take into account correlated noise between fMRI voxels better recover the amount of uncertainty (quantified as the width of a probability distribution over possible stimuli) associated with the decoded estimate. Furthermore, we show that not all correlated variability should be treated equally, as modeling tuning-dependent correlations has the greatest impact on decoding performance. Next, we examine actual noise correlations in human visual cortex, and find that shared variability in areas V1-V3 depends on the tuning properties of fMRI voxels. In line with our simulations, accounting for this shared noise between similarly tuned voxels produces important benefits in decoding. Our findings underscore the importance of accurate noise models in fMRI decoding approaches, and suggest a statistically feasible method to incorporate the most relevant forms of shared noise.

# 1 Introduction

What sensory stimulus evoked this particular pattern of cortical activity? This question lies at the heart of brain decoding algorithms. Most fMRI decoders will estimate or 'decode' a single stimulus value that is, according to some underlying model, most consistent with the observed data. In truth, however, there is rarely just one stimulus that provides a plausible explanation. Rather, different stimuli may all be somewhat consistent with the measured response. The main reason for this imprecision is variability (or noise), which causes even the same stimulus to elicit different activity patterns each time the stimulus is presented. By the same token, variability allows the same response pattern to be evoked by a range of different stimuli. From a noisy response pattern, therefore, the stimulus that elicited the pattern cannot be inferred with perfect precision. Rather, there is some degree of *uncertainty* in the predictions, and this uncertainty may vary from one decoded activity pattern to the next. Importantly, while uncertainty may stem from imprecise measurements, it can also be of neural origin. Since neural responses themselves are inherently noisy (Dean, 1981; Schiller et al., 1976), cortical activity cannot encode stimulus information with perfect precision, and this imprecision can, moreover, fluctuate over time. Uncertainty, thus, is an important feature of cortical stimulus representations, providing a window on the fidelity of neural stimulus processing from one moment to the next.

How can uncertainty be measured from cortical activity patterns? Mathematically speaking, the presence of uncertainty means that cortical information is most accurately characterized by a probability distribution over all possible stimuli. The wider this distribution, the larger is the range of stimuli that could have evoked the observed pattern of cortical activity, and hence, the higher is the uncertainty. In order to measure uncertainty, therefore, we should estimate probability distributions. But how can this be achieved, and why is this not possible with conventional decoding algorithms? Recall that uncertainty largely stems from noise in the data. This noise turns the causal relationship between stimuli and responses from a deterministic to a stochastic one, described by probabilities rather than fixed outcomes. To estimate probabilities, therefore, a decoding algorithm should capture this noisy, stochastic relationship. Mathematical models that describe the causal link between stimuli and cortical activity are typically known as *forward* or *generative models*, and have become a popular tool to describe (and extract information from) fMRI activity. Importantly, however, most models to date assume that noise is simply independent between fMRI voxels (e.g. Kay, Naselaris, Prenger, & Gallant, 2008; Brouwer & Heeger, 2009; Serences, Saproo, Scolari, Ho, & Muftuler, 2009; Jehee, Ling, Swisher, van Bergen, & Tong, 2012; Ester, Anderson, Serences, & Awh, 2013). Contrary to this assumption, mounting evidence suggests that variability is instead correlated in cortex (Arcaro et al., 2015; Bair et al., 2001; de Zwart et al., 2008; Henriksson et al., 2015; Parkes et al., 2005; Rosenbaum et al., 2016; Smith and Kohn, 2008; Zohary et al., 1994). It is well known that given incorrect assumptions of independence, decoding algorithms may fail to fully characterize the probability distribution encoded in cortical activity, despite producing reasonable estimates of the most likely presented stimulus (Domingos and Pazzani, 1997; Niculescu-Mizil and Caruana, 2005; Zhang, 2004).

Thus, the ability to measure stimulus distributions from cortical activity patterns hinges on having an appropriate model of the noise correlations in the data. But since the number of these correlations increases quadratically with the number of voxels in an fMRI data set, estimating them individually and without any guiding principles is often statistically impossible. Here, we propose a simpler approach, based on the notion that not all shared noise is equally important. Specifically, as others have argued before (Abbott and Dayan, 1999; Averbeck and Lee, 2006; Moreno-Bote et al., 2014; Smith and Kohn, 2008), correlated noise is most detrimental when it is indistinguishable from the stimulus-driven response. That is, when noise is correlated between similarly-tuned voxels, their joint activation can either indicate the presence of a mutually preferred stimulus, or that of shared noise. A decoder ignorant of the possibility of such correlated noise would tend to conclude that the voxels were activated by their preferred stimulus. Accordingly, we reasoned, this "naïve" decoder would incorrectly assign high probabilities to the stimuli preferred by these voxels, without considering interpretations consistent with shared noise.

To quantify these intuitions, we will first demonstrate in simulations that an accurate characterization of probability distributions is possible if, specifically, those correlations are accounted for that align with similarities in voxel tuning preferences. Correlations that do not have such tuning-dependent structure, on the other hand, may be safely ignored. We then examine noise correlations in fMRI measurements from human visual cortex, and find that these correlations contain the relevant tuning-related structure. Finally, we show that a decoding model that takes these tuning-dependent noise correlations into account provides an accurate window onto trial-by-trial fluctuations in the uncertainty in cortical stimulus representations. These findings exemplify the importance of incorporating noise correlations in forward models of neuroimaging data, and suggest a simple, statistically feasible approach to do so.

## 2 Methods

### 2.1 Participants
Eighteen healthy adult volunteers (aged 22-31 years, seven female) participated in this study. All had normal or corrected-to-normal vision, and provided written and informed consent prior to participating. The study was approved by the Radboud University Institutional Review Board.

### 2.2 MRI data acquisition
MRI data were acquired using a Siemens 3T Magnetom Trio scanner with an eight-channel occipital receiver coil, located at the Donders Center for Cognitive Neuroimaging. At the start of each session, a high-resolution T1-weighted magnetization-prepared rapid gradient echo anatomical scan (MPRAGE, FOV 256 x 256, 1 mm isotropic voxels) was acquired for each participant. Functional imaging data were collected in 30 slices oriented perpendicular to the calcarine sulcus, covering all of the occipital and part of posterior parietal and temporal cortex, using a T2*-weighted gradient-echo echoplanar imaging sequence (TR 2000 ms, TE 30 ms, flip angle 90°, FOV 64 x 64, slice thickness 2.2 mm, in-plane resolution 2.2 x 2.2 mm). Data were previously analyzed in the context of a different study (van Bergen et al., 2015).

### 2.3 Experimental design & stimuli
Visual stimuli were displayed on a rear-projection screen using a luminance-calibrated EIKI projector (resolution 1024 x 768 pixels, refresh rate 60 Hz). Observers viewed the screen through a mirror mounted on the head coil. Stimuli were generated using a Macbook Pro computer running Matlab and the Psychophysics Toolbox (Brainard, 1997; Pelli, 1997).

Throughout each experimental run, participants were required to maintain fixation at a bull's eye target (radius: 0.25°) that was shown at the center of the screen. Each run started with an initial fixation period (4 s), followed by 18 stimulus trials (12 s each, separated by inter-trial intervals of 4 s) and a final fixation period (4 s). At the start of each trial, an orientation stimulus was shown, consisting of a counterphasing sinusoidal grating (duration: 1.5 s, contrast: 10%, spatial frequency: 1 cycle/°, randomized spatial phase, 2 Hz sinusoidal contrast modulation), and presented in an annular aperture surrounding fixation (inner radius: 1.5°, outer radius: 7.5°, grating contrast decreased linearly to 0 over the outer and inner 0.5° radius of the annulus). Stimulus orientations were determined pseudo-randomly (from 0-179°) to ensure an approximately even distribution of orientations in each run. Shortly (6.5 s) after the stimulus had disappeared, a black line (length: 2.8°, width: 0.1°) appeared at the center of the screen at an initially random orientation. The line remained on screen for 4 s, and disappeared gradually over the last 1 s of this window. Participants gave their response by adjusting the orientation of this line to match the previously seen grating, using separate buttons for clockwise or counterclockwise rotation on an MRI-compatible button box.

Participants completed 10-18 stimulus runs. Each scan session also included two runs of visual localizer stimuli. In these runs, flickering checkerboard patterns (check size: 0.5°, contrast: 100%, display rate: 10 Hz, i.e. a new random checkerboard pattern was presented every 100 ms) were shown in 12 second blocks, interleaved with fixation blocks of equal duration. Checkerboards were presented in the same visual aperture as the orientation gratings.

Retinotopic maps of visual cortex were acquired in a separate scan session using standard retinotopic mapping procedures (DeYoe et al., 1996; Engel et al., 1997; Sereno et al., 1995).

## 2.4 Functional MRI data preprocessing and regions of interest

Functional images were motion-corrected using FSL's MCFLIRT (Jenkinson et al., 2002), and aligned to a previously collected anatomical reference scan using FreeSurfer (Fischl et al., 1999). To remove slow drifts in the BOLD signal, voxel time series were temporally filtered with a high-pass cut-off period of 40 seconds. No slice timing correction was performed. Residual variations in the BOLD signal induced by motion were removed through linear regression (using 18 motion regressors, consisting of the 6 translation & rotation estimates from MCFLIRT, the squares of these numbers and their temporal derivatives).

Regions of interest (ROIs; V1, V2 and V3) were defined on the reconstructed cortical surface using conventional procedures (DeYoe et al., 1996; Engel et al., 1997; Sereno et al., 1995). Within each retinotopically defined area, all voxels that responded to the localizer stimulus above a lenient threshold ($p < 0.05$ uncorrected) were selected for subsequent analysis. Functional time series data were analyzed in the native space for each participant.

Voxel time series were Z-normalized using corresponding time points of all trials within each run (that is, activity in the $N^{th}$ time point within each trial was normalized by the mean and standard deviation of activity across the $N^{th}$ time points of all trials within the same run). To obtain an activation pattern for each trial, activity was averaged across the first 4 s of each trial, after shifting each time series by 4 seconds to account for hemodynamic lag in the BOLD signal. This choice of time window ensured that no activity from the subsequent response window (which started 8 s after trial onset) was included in the analyses.

## 2.5 Simulation procedures

To determine how noise correlations impact decoding performance, we simulated fMRI data with known covariance structure, and tested the degree to which each of four decoding models recovered the information available in the simulated fMRI activation patterns. Specifically, for each of 10 hypothetical observers, we generated fMRI activation patterns for each of 1000 visual orientation stimuli, using $M = 500$ voxels. The response of a single voxel on a given trial was simulated as:

$$b_i = f_i(s) + \varepsilon_i \qquad (1)$$

where $b_i$ is the response of the $i$-th voxel, $s$ is the orientation of the simulated stimulus (drawn from a uniform distribution over the interval $[0, 180)°$), $f_i(s)$ is voxel $i$'s orientation tuning curve and $\varepsilon_i$ is random noise (see also **Fig. 1A**). Across voxels, this results in a vector (or "pattern") of activity $\mathbf{b} = \{b_i\}$. The noise across voxels $\boldsymbol{\varepsilon} = \{\varepsilon_i\}$ was drawn from a multivariate Normal distribution with mean $\mathbf{0}$ and voxel-by-voxel covariance $\boldsymbol{\Sigma}$:

$$\boldsymbol{\varepsilon} \sim \mathcal{N}(\mathbf{0}, \boldsymbol{\Sigma}) \tag{2}$$

The noise covariance matrix $\boldsymbol{\Sigma}$ was simulated as a combination of a correlation matrix $\mathbf{R}^{(full)} = \{R_{ij}^{(\text{full})}\}$ and a vector of standard deviations $\boldsymbol{\tau} = \{\tau_i\}$, such that the covariance between a pair of voxels $(i,j)$ was given by:

$$\Sigma_{ij} = \tau_i \tau_j R_{ij}^{(\text{full})} \tag{3}$$

Each element of the vector $\boldsymbol{\tau}$ was drawn randomly from a Normal distribution with mean 3 and variance $0.2^2$, and the noise correlation $R_{ij}^{(\text{full})}$ between voxels $i$ and $j$ was given by:

$$R_{ij}^{(\text{full})} = \begin{cases} 1, & \text{if } i = j \\ R_{ij}^{(\text{tuning})} + R_{ij}^{(\text{arbitrary})}, & \text{if } i \neq j \end{cases} \tag{4}$$

in which $R_{ij}^{(\text{tuning})}$ is a function of the tuning similarity between voxels $i$ and $j$, defined as the correlation between their tuning curves:

$$R_{ij}^{(\text{tuning})} = \begin{cases} 1, & \text{if } i = j \\ 0.2 \times \text{corr}\left(f_i(s), f_j(s)\right), & \text{if } i \neq j \end{cases} \tag{5}$$

and $\mathbf{R}^{(arbitrary)} = \{R_{ij}^{(arbitrary)}\}$ was created by shuffling the matrix $\mathbf{R}^{(tuning)} = \{R_{ij}^{(\text{tuning})}\}$, such that its rows and columns were rearranged in the same, random order (which leaves the diagonal of the matrix intact). That is, if $\mathbf{z}$ is a random re-ordering of voxel indices $1:M$, then the arbitrary noise correlation between voxels $i$ and $j$ is given by:

$$R_{ij}^{(arbitrary)} = R_{z_i z_j}^{(\text{tuning})} \tag{6}$$

This procedure alters the structure of noise correlations such that they become independent of tuning similarity, while keeping overall levels of correlated noise constant, allowing for an appropriate comparison between the effects of the two correlation structures, $\mathbf{R}^{(tuning)}$ and $\mathbf{R}^{(arbitrary)}$, on decoding performance.

Orientation tuning curves $\boldsymbol{f}(s) = \{f_i(s)\}$ were simulated as a linear combination of 8 bell-shaped basis functions:

$$f_i(s) = \sum_{k=1}^{8} W_{ik} g_k(s) \tag{7}$$

where $g_k(s)$ is the $k$-th basis function, and $W_{ik}$ is the weighting coefficient on this basis function for the $i$-th voxel. Basis functions were positive half-wave rectified cosine functions raised to the fifth power (cf. Brouwer & Heeger, 2011; van Bergen, Ma, Pratte, & Jehee, 2015):

$$g_k(s) = \left[\cos\left(\frac{\pi}{90}(s - \varphi_k)\right)\right]_+^5 \tag{8}$$

where $\varphi_k$ is the preferred orientation of the $k$-th basis function. Preferred orientations were equally spaced, with one basis function tuned maximally towards horizontal orientations. Weights were drawn randomly from a standard Normal distribution.

Given these simulation procedures, the conditional probability of a voxel activation pattern given a stimulus is defined as:

$$p(\mathbf{b}|s) \propto \exp\left((\mathbf{b} - \mathbf{f}(s))^T \Sigma^{-1}(\mathbf{b} - \mathbf{f}(s))\right) \tag{9}$$

## 2.6 Decoding algorithm (simulations)

To test what sources of correlated noise are particularly relevant to decoding performance, we compared three decoding models, each of which ignored some aspects of the simulated data. Specifically, we assumed that the three decoders had perfect knowledge of voxel tuning functions $\mathbf{f}(s) = \{f_i(s)\}$, but varied in their assumed structure of voxel covariances $\Sigma$. That is, the covariance between two voxels is given by:

$$\Sigma_{ij} = \tau_i \tau_j R_{ij} \tag{10}$$

While noise standard deviations $\tau_i$ and $\tau_j$ of voxels $i$ and $j$ were kept constant across the decoding models, each of the decoders made different assumptions regarding the correlation matrix $\mathbf{R} = \{R_{ij}\}$. The first decoder assumed that $\mathbf{R}^{(naïve)}$ is the $M \times M$ identity matrix that captures no shared noise between voxels. The second decoder with correlation matrix $\mathbf{R}^{(arbitrary)}$ accounted for shared variability between voxels of arbitrary structure and the third decoder captured tuning-dependent correlations $\mathbf{R}^{(tuning)}$.

By applying Bayes rule with a flat stimulus prior, we obtained for each simulated trial of voxel activity the posterior probability distribution over stimulus orientation:

$$p(s|\mathbf{b}; \mathbf{R}) = \frac{p(\mathbf{b}|s; \mathbf{R})p(s)}{\int p(\mathbf{b}|s; \mathbf{R})p(s)\,ds} \tag{11}$$

where $\mathbf{R}$ indicates the noise model that was used for decoding. Thus, the actual posterior distribution encoded in a pattern of activity is given by $p(s|\mathbf{b}; \mathbf{R}^{(full)})$, while the posteriors decoded using alternative models are denoted by $p(s|\mathbf{b}; \mathbf{R}^{(naïve)})$, $p(s|\mathbf{b}; \mathbf{R}^{(arbitrary)})$ and $p(s|\mathbf{b}; \mathbf{R}^{(tuning)})$. The circular mean of each decoded posterior distribution served as the orientation point-estimate, while its width (circular standard deviation) quantified the degree of uncertainty in that estimate (see **Fig. 1A**). These statistics were computed by numerical integration.

## 2.7 Decoding benchmarks (simulations)

We used three benchmark tests to evaluate and compare between the three decoders. The first test focused on decoding accuracy of the presented orientations. Decoding accuracy was quantified by computing the circular correlation coefficient between the presented and decoded orientations across trials, for each noise model and each simulated observer (see **Fig. 1C**). The second test focused on the degree of uncertainty in the decoded orientation estimates. Specifically, for each decoder and simulated observer, we took the width (or uncertainty) of the decoded posterior distribution for each simulated trial of data, and correlated this with the actual uncertainty in the data (that is, the uncertainty in $p(s|\mathbf{b}; \mathbf{R}^{(full)})$) for the same set of data (see **Fig. 1D**). The third test assessed the overall amount of information lost when a posterior distribution is decoded using an incomplete noise model. This information loss was quantified as the Kullback-Leibler divergence from each decoded distribution ($p(s|\mathbf{b}; \mathbf{R}^{(naïve)})$, $p(s|\mathbf{b}; \mathbf{R}^{(arbitrary)})$ and $p(s|\mathbf{b}; \mathbf{R}^{(tuning)})$) to the true posterior distribution ($p(s|\mathbf{b}; \mathbf{R}^{(full)})$), for each trial of simulated data (see **Fig. 1E**). The KL-divergence from a distribution $Q$ to a target distribution $P$ is defined as:

$$D_{\mathrm{KL}}(P \parallel Q) = \int p(x) \log \frac{p(x)}{q(x)} \mathrm{d}x \tag{12}$$

KL-divergences were calculated numerically for each noise model and each trial, and then averaged across trials.

To assess whether the correlations computed in the first two benchmark tests were significantly different between the three decoders, we ran paired t-tests across observers on the Fisher-transformed correlation coefficients. The mean correlation coefficient across observers was calculated by taking the mean of the Fisher-transformed values, and then transforming this mean back to the correlation scale. Differences in KL-divergence between the three decoders were evaluated using paired t-tests across observers, on the mean KL-divergences for each noise model across trials.

## 2.8 Analysis of fMRI data

*2.8.1 Tuning & noise.* Voxel tuning and noise properties were estimated based on equation (1). Specifically, we assumed that the response of each voxel on a given trial was a sum of an orientation tuning function and multivariate Normally distributed random noise (equation (2)). Each voxel tuning function was assumed to be a linear combination of the idealized, bell-shaped tuning functions of eight hypothetical neural populations. Thus, the tuning properties of each voxel $i$ are summarized by the coefficients $\mathbf{W}_i$ on these eight basis functions, and estimating a voxel's tuning properties consisted in estimating these coefficients.

The vector (or pattern) of voxel responses measured on trial $t$ is denoted by $\mathbf{b}^{(t)}$, and $\mathbf{B} = [\mathbf{b}^{(1)}, ..., \mathbf{b}^{(N)}]$ is the $M \times N$ matrix of all such patterns for a given observer (where $M$ is the number of voxels and $N$ the number of trials for that observer). Tuning coefficients $\mathbf{W}$ were estimated using ordinary least squares (OLS) regression:

$$\widehat{\mathbf{W}} = \mathbf{B} g(\mathbf{s})^{\mathrm{T}} (g(\mathbf{s}) g(\mathbf{s})^{\mathrm{T}})^{-1} \tag{13}$$

where $g(\mathbf{s}) = \{g_k(s_t)\}$ are the predicted responses of the eight neural populations to the sequence of presented stimuli across trials (see equation (8)). Control analyses verified that our results were robust to the particular choice of basis functions used to describe voxel tuning curves (**Supplementary fig. 1**). Note that if correlated noise were known to depend on tuning similarities between voxels, an OLS regression would not be the most statistically efficient estimator. However, since we do not want to assume *a priori* that tuning and noise correlations are linked (which would constitute circular inference), the OLS estimator is appropriate for these analyses.

Next, we computed the tuning similarity between voxels by calculating the correlation coefficient between their tuning curves. To ensure that the estimated correlation coefficient did not reflect noise shared between same-trial voxel responses, tuning similarity between each pair of voxels was calculated based on asynchronous presentations of orientation stimuli. To this end, we split our fMRI data into two independent sets. The first set consisted of voxel activations and stimuli $\{\mathbf{B}^{(1)}, \mathbf{s}^{(1)}\}$ from odd-numbered fMRI runs, while the second set $\{\mathbf{B}^{(2)}, \mathbf{s}^{(2)}\}$ came from even-numbered fMRI runs. By running the OLS regression separately for each set of data, we obtained two sets of estimated tuning coefficients, $\widehat{\mathbf{W}}^{(1)}$ and $\widehat{\mathbf{W}}^{(2)}$. For each pair of voxels $(i,j)$, we then determined the similarity in their tuning preferences by cross-correlating the tuning curves estimated on the two sets of data:

$$\widehat{r_{tuning}}(i,j) = \text{corr}\left(\widehat{\mathbf{W}}_i^{(1)} \boldsymbol{g}(\mathbf{s}), \widehat{\mathbf{W}}_j^{(2)} \boldsymbol{g}(\mathbf{s})\right) \tag{14}$$

The level of noise on a given trial was computed as the difference between a voxel's response and the value predicted by its estimated tuning for the presented stimulus:

$$\widehat{\varepsilon}_{it} = \begin{cases} b_i^{(t)} - \widehat{\mathbf{W}}_i^{(1)} \boldsymbol{g}(s_t), & \text{if } t \in \text{partition 1} \\ b_i^{(t)} - \widehat{\mathbf{W}}_i^{(2)} \boldsymbol{g}(s_t), & \text{if } t \in \text{partition 2} \end{cases} \tag{15}$$

Finally, the noise correlation between voxels $i$ and $j$ was estimated as:

$$\widehat{r_{noise}}(i,j) = \text{corr}(\widehat{\boldsymbol{\varepsilon}}_i, \widehat{\boldsymbol{\varepsilon}}_j) \tag{16}$$

*2.8.2 Function fits.* To determine how noise correlations might depend on the degree of similarity in tuning preferences between voxels, we first sorted pairs of voxels into 20 bins of similar tuning correlations. Within each bin, we then calculated the average Fisher-transformed noise correlation and tuning correlation, across all voxel pairs in the bin. To these bin-averages, we subsequently fit the following exponential decay function:

$$h(n) = \alpha \exp\left(-\beta\left(1 - \langle\widehat{r_{tuning}}\rangle_n\right)\right) + \gamma \tag{17}$$

where $\langle\widehat{r_{tuning}}\rangle_n$ is the mean (estimated) tuning similarity between voxel pairs in the $n$-th bin. This function describes a decay in noise correlations with decreasing tuning similarity between voxels. This decay starts from an initial value set by parameter $\alpha$, and the rate of decay is governed by $\beta$. The third parameter $\gamma$ describes an overall baseline correlation between voxels. These parameters were fit by minimizing the sum of squared residuals $S$, between the measured and predicted (Fisher-transformed) noise correlations:

$$S = \sum_n \left(\langle\text{arctanh}(\widehat{r_{noise}})\rangle_n - \text{arctanh}(h(n))\right)^2 \tag{18}$$

where $\langle\text{arctanh}(\widehat{r_{noise}})\rangle_n$ is the mean Fisher-transformed (estimated) noise correlation between voxel pairs in the $n$-th bin. Parameter fits were constrained such that the amplitude ($\alpha$) and decay rate ($\beta$) were always non-negative. To assess the goodness-of-fit of the exponential decay function to the data, we computed the adjusted coefficient of determination ($R^2_{adj}$), which includes a correction for the degrees of freedom in each model. A one-tailed Wilcoxon-signed rank test was then used to determine whether the $R^2_{adj}$ values across participants were reliably larger than 0, which would indicate a significant amount of variance explained (note that a one-tailed test is appropriate since our one-sided hypothesis is that the function explains more variance than expected by chance, not less).

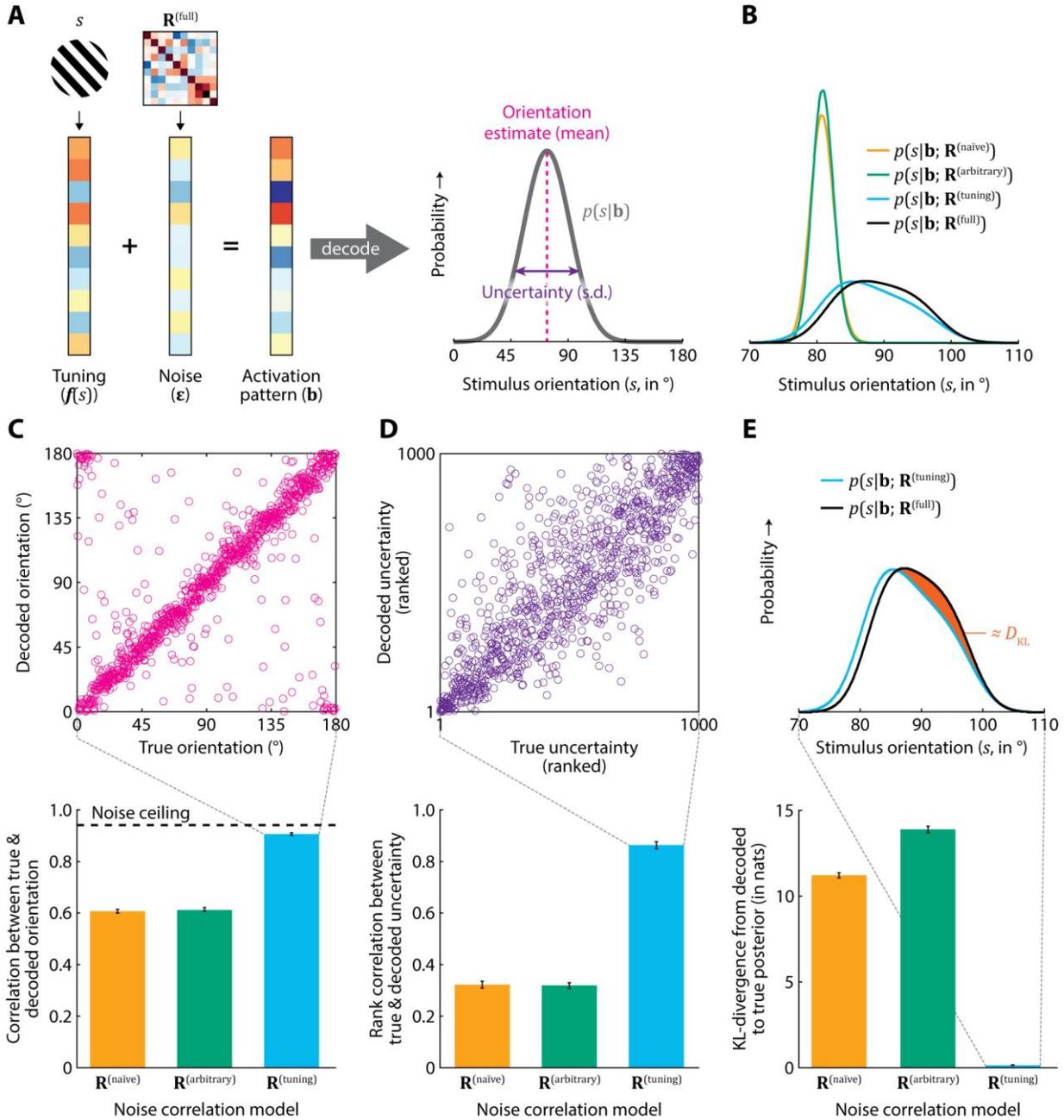

**Figure 1:** Simulations to compare the effect of noise correlations on decoding performance. (**A**) Illustration of simulation procedures and decoder output. Each activation pattern (**b**) is simulated as a sum of a stimulus-dependent (tuning) component and random noise. This noise has a voxel-by-voxel correlation structure $\mathbf{R}^{(\text{full})}$. Given a generative model of the voxel tuning properties and noise correlations, a posterior distribution $p(s|\mathbf{b})$ can be calculated. The (circular) mean of this distribution serves as the point estimate of the simulated stimulus orientation, while its width (circular standard deviation) quantifies the amount of uncertainty in this estimate. (**B**) Posterior distributions calculated with different noise correlation models, for an example trial of simulated data. Note that the distribution computed using the $\mathbf{R}^{(\text{tuning})}$ model (shown in blue) is very close to the true posterior distribution (shown in black). The distribution calculated using arbitrary noise correlations (shown in green) is much less accurate, and very similar to the posterior distribution obtained with the naïve decoder (shown in yellow). (**C**) Top: orientation estimates for the $\mathbf{R}^{(\text{tuning})}$ decoder, for one hypothetical observer. Bottom: comparison of point-estimate orientation decoding accuracy between decoders, quantified as the circular correlation between presented (simulated) and decoded orientations. The noise ceiling was calculated as the accuracy achieved by a decoder that has full knowledge of all noise correlations ($\mathbf{R}^{(\text{full})}$).

# 3 Results

This paper examines the relevance of shared noise to the decoding of stimulus information from cortical activity. Specifically, we will contrast two forms of shared noise: noise that is shared between voxels similarly tuned to the decoded stimulus feature, and noise that is correlated but does not depend on voxel tuning preference (i.e., it has arbitrary structure). We first provide a theoretical comparison of these two forms of noise, using simulations, before turning to an investigation of the actual structure of fMRI noise correlations in human visual cortex, and the degree to which shared noise affects decoding accuracy in practice.

## 3.1 Decoded information is dominated by tuning-dependent noise correlations

We will first demonstrate how noise correlations can impact decoding performance in theory. In order to do so, we simulated fMRI data for 10 hypothetical observers, who each participated in 1000 experimental trials in which orientation stimuli were shown. Voxel responses on each trial were simulated as the sum of an orientation tuning curve and random noise (**Fig. 1A**, see also Methods). Some of this noise was independent and specific to each individual voxel, while the remainder was shared between voxels. Specifically, we introduced noise that is shared between voxels of similar orientation tuning preference, and noise shared randomly between voxels irrespective of their tuning properties. Each of these two forms of shared noise had an equal contribution to the overall levels of noise in the data.

Each pattern of simulated data contains information about the stimulus that was presented to the hypothetical observer on that trial. Because the data is noisy, the amount of information contained in the pattern of activity is necessarily limited: even with perfect knowledge of the simulation parameters, the stimulus cannot be inferred with absolute accuracy. Rather, there is some degree of *uncertainty* about the stimulus that gave rise to the observed activation pattern. Mathematically, the information contained in a pattern of activity can be described by the posterior distribution $p(s|\mathbf{b})$, which gives the probability that stimulus $s$ was presented to the hypothetical observer, given the observed activation pattern $\mathbf{b}$ (**Fig. 1A** – note that the posterior also reflects prior knowledge, but we assume an uninformative prior distribution here (see Methods)). The wider this distribution, the greater the range of stimuli consistent with the observed pattern of voxel activity, and the larger the uncertainty about $s$. Importantly, this posterior distribution depends on the assumptions used in the decoder, and the degree to which these assumptions match the generative model by which the data arose (i.e., the model and parameter values

---

(**D**) Top: uncertainty estimates for the $\mathbf{R}^{(\text{tuning})}$ decoder, for one simulated observer. The true uncertainty in the data was computed as the circular standard deviation of the posterior distribution obtained using a decoder that has full knowledge of all noise correlations ($p(s|\mathbf{b}; \mathbf{R}^{(\text{full})})$), for each trial of data. Plotted are the ranks of the true and decoded uncertainty values within the 1000 trials in the data set. Bottom: comparison of uncertainty decoding accuracy across decoders, quantified as the rank correlation (Spearman's rho) between true and decoded uncertainty. (**E**) Top: illustration of Kullback-Leibler divergence ($D_{KL}$) to the true posterior distribution, from a posterior distribution obtained with the $\mathbf{R}^{(\text{tuning})}$ model, for one example trial (the same trial shown in **B**). The orange-shaded region corresponds to the area between the distributions that contributes positively to the KL-divergence. Bottom: comparison of the overall accuracy of posterior distributions produced by each decoder, quantified as the KL-divergence from the decoded to the true posterior distribution. This KL-divergence measures the information lost by approximating the true posterior with a posterior distribution decoded under different assumptions. In **C-E**, bars and error bars correspond to the mean +/- 1 SEM.

used in the simulation). Here, we will evaluate which assumptions are more, and which are less, important for recovering the information contained in the data.

Our focus is on assumptions regarding noise; accordingly, we assume that the decoders compared here have perfect knowledge of the simulated voxel tuning properties. We denote by $p(s|\mathbf{b}; \mathbf{R}^{(\text{full})})$ the posterior probability distribution of a decoder that has full knowledge of the simulated noise correlation structure $\mathbf{R}^{(\text{full})}$. How accurately can $p(s|\mathbf{b}; \mathbf{R}^{(\text{full})})$ be approximated if some or all of the noise correlations in $\mathbf{R}^{(\text{full})}$ are ignored? To answer this question, we will compare the performance of three decoders, each of which uses different assumptions regarding the noise correlation structure. The first decoder, with (diagonal) noise correlation structure $\mathbf{R}^{(\text{naïve})}$, is a naïve decoder that assumes independent noise between voxels (i.e., no noise correlations). The second decoder only has knowledge of shared noise that is unrelated to voxel tuning properties, with correlation structure $\mathbf{R}^{(\text{arbitrary})}$. The third decoder only accounts for those noise correlations that align with voxel tuning similarities, with structure $\mathbf{R}^{(\text{tuning})}$. Using each of these decoders, we calculated a posterior distribution over stimulus orientation for each trial of simulated voxel activity, and compared the so-obtained posterior distributions with $p(s|\mathbf{b}; \mathbf{R}^{(\text{full})})$.

An example of the posterior distributions obtained using each of the three decoders for one simulated trial of data is shown in **Fig. 1B**. Also shown in this figure is the posterior distribution that was obtained using full knowledge of the simulation's parameter values (i.e., $p(s|\mathbf{b}; \mathbf{R}^{(\text{full})})$), which describes the actual information contained in the activity pattern. As can be seen in the figure, a decoder that uses naïve assumptions regarding the covariance structure produces a posterior distribution that is very different from the true posterior distribution $p(s|\mathbf{b}; \mathbf{R}^{(\text{full})})$, concentrating probability in a narrow region. Interestingly, the decoder that accounts for shared voxel noise that is of arbitrary structure performs about as poorly as the naïve decoder, producing a nearly identical and similarly overconfident posterior distribution. Incorporating tuning-dependent correlations into the decoder, however, makes a large difference in terms of performance: the decoded posterior distribution of this decoder very closely approximates the true posterior distribution contained in the simulated pattern of activity on that trial.

To quantify and summarize these observations across simulated trials and subjects, we relied on three benchmark tests. For the first test, we compared the accuracy of the orientation estimates produced by each decoder, quantified as the circular correlation coefficient between simulated and decoded orientations (**Fig. 1C**). For the naïve decoder, the orientation estimates were fairly accurate ($\langle r \rangle$ = 0.61; one-sample t-test: $t(9)$ = 59.21, $p < 10^{-12}$). Compared to the naïve decoder, a decoder that assumes arbitrary correlations between voxels did not significantly increase in decoding performance (paired t-test; $t(9)$ = 2.06, $p$ = 0.07). Accounting for tuning-dependent correlations, however, caused a substantial and significant improvement with respect to both the naïve and arbitrary-shared-noise decoders (paired t-tests; $t(9)$ = 25.04, $p < 10^{-8}$ & $t(9)$ = 25.64, $p < 10^{-8}$, respectively). Thus, a decoder that captures tuning-dependent noise correlations outperforms the other two decoders in terms of its estimates of the presented stimulus orientation.

The second benchmark test examined how well each decoder captured the amount of uncertainty in the decoded stimulus orientations. Specifically, we assessed the degree to which the width of the decoded posterior distributions matched that of the true posterior distribution $p(s|\mathbf{b}; \mathbf{R}^{(\text{full})})$, for each of the three decoders and across trials (**Fig. 1D**). This was done by computing the rank correlation coefficient (Spearman's rho) between the true and decoded uncertainty values. Although significant, this correlation for the naïve decoder was rather weak ($\langle r \rangle$ = 0.32; one-sample t-test: $t(9)$ = 22.36, $p < 10^{-8}$). Compared to the naïve decoder, accounting for arbitrary noise correlations did not significantly improve decoding accuracy of uncertainty (paired t-test; $t(9) = -0.85$, $p = 0.42$). The decoder that captures tuning-dependent noise correlations, however, again outperformed the other two noise models on this test (paired t-tests; $t(9) = 21.31$, $p < 10^{-8}$ & $t(9) = 20.58$, $p < 10^{-8}$, for the comparison between naïve and tuning-dependent decoders, and arbitrary and tuning-dependent decoders, respectively). This advantage of the tuning-dependent decoder was rather pronounced, with uncertainty decoding accuracy increasing from $r \approx 0.3$ to $r \approx 0.8$ when tuning-dependent shared noise was accounted for.

The third and final test computed the Kullback-Leibler divergence from the decoded posterior distribution to the true posterior distribution, for each of the three decoders and on every trial (**Fig. 1E**). The KL-divergence measures the information that is lost when the true posterior is approximated by an alternative distribution. Interestingly, this loss is nearly zero for the decoder that accounts for tuning-dependent noise correlations – despite the fact that this decoder ignores roughly half of all shared noise in the data. By comparison, for the naïve and arbitrary-correlations decoders, the decoded posterior distributions diverge much more strongly from the true posterior distribution (paired t-tests vs. the $\mathbf{R}^{(tuning)}$ model; $t(9) = 77.06$, $p < 10^{-13}$ & $t(9) = 79.98$, $p < 10^{-13}$).

Together, these results indicate that tuning-dependent noise correlations have the greatest impact on the amount of information that can be recovered from patterns of cortical activity – much more so than other forms of correlated noise. Although the actual amount of noise was equally large for either arbitrary or tuning-dependent correlations, a decoder that accounted for arbitrary correlations had almost no advantage in terms of decoding accuracy on a naïve decoder. Rather, it was the decoder that captured tuning-dependent correlations that showed the greatest improvement in decoding accuracy of the posterior distributions contained in the simulated data.

### 3.2 Noise correlations in human visual cortex

Our simulation results demonstrate that tuning-dependent noise correlations are especially relevant to decoding performance, while other correlations may be safely ignored. But does noise in actual fMRI data depend on tuning similarity between voxels? To investigate this, we re-analyzed fMRI data obtained from 18 human participants, while they viewed oriented grating stimuli (in the context of a different study (van

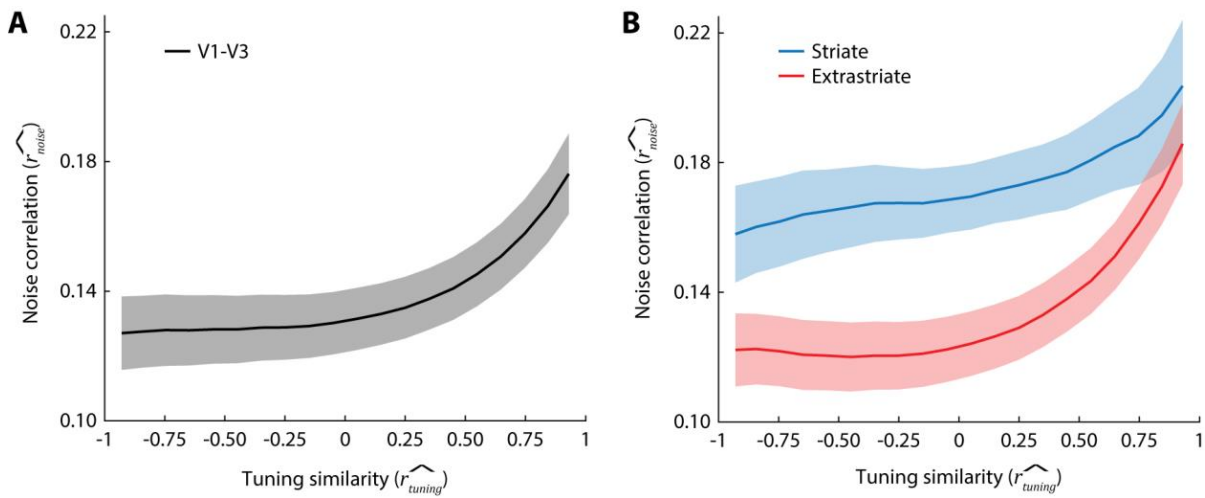

**Figure 2:** Average relationship between $\widehat{r_{noise}}$ and $\widehat{r_{tuning}}$, across 18 observers, in V1-V3 combined (**A**) and separately for striate and extrastriate areas (**B**). Results for individual subjects are shown in **Supplementary fig. 2**. Lines and shaded regions show the mean +/- 1 SEM noise correlation across observers, in 20 equally spaced tuning similarity bins.

Bergen et al., 2015)). We analyzed cortical activity evoked by each stimulus in visual areas V1, and V2 and V3 combined. The fMRI responses were used to estimate a tuning curve for each voxel, and noise on each trial was calculated by subtracting the value predicted by the voxel's tuning curve from the voxel's response. For each pair of voxels, we then computed the similarity in the estimated tuning curves ($\widehat{r_{tuning}}$) and the degree of noise shared between the two voxels ($\widehat{r_{noise}}$). Voxel pairs were subsequently sorted into bins of similar tuning correlations, and averages were computed across the data in each bin (**Fig. 2**). To determine whether the degree of shared noise was related to similarities in voxel tuning curves, we fit the bin averages with an exponential decay function. The goodness-of-fit of this function was quantified by computing for each observer the overfitting-adjusted coefficient of determination ($R^2_{adj}$), and a Wilcoxon signed-rank test on the $R^2_{adj}$ values was used to assess whether the exponential decay function explained significant variance in the data, across observers.

The average relationship between $\widehat{r_{tuning}}$ and $\widehat{r_{noise}}$ across visual areas V1, V2 and V3 is shown in **Fig. 2A**. As can be seen in this figure, noise correlations appear to be largest between voxels that have very similar tuning properties ($\widehat{r_{tuning}}$ near 1), and then diminish as tuning curves grow more dissimilar. An exponential decay with decreasing tuning similarity described this relationship very well across participants ($\langle R^2_{adj}\rangle$ = 0.73, *p* < 0.001). When voxels are split up into striate (V1) and extrastriate areas (V2 & V3), the average relationship between tuning similarity and shared noise (shown in **Fig. 2B**) in these ROIs is similar to that in the combined ROI. In both ROIs, the fitted exponential decay functions explained significant variance across observers (striate ROI: $\langle R^2_{adj}\rangle$ = 0.45, *p* < 0.01; extrastriate ROI: $\langle R^2_{adj}\rangle$ = 0.83, *p* < 0.001). This finding was robust to the particular choice of functions used to fit voxel tuning curves (**Supplementary fig. 1**). These results indicate that tuning-dependent shared noise, which our simulations revealed to be so important for decoding, is present in fMRI data from human visual cortex.

Neighboring voxels in cortex tend to have similar tuning properties (Freeman et al., 2011; Mannion et al., 2010; Sasaki et al., 2006; Swisher et al., 2010). At the same time, various nuisance variables, such as eye or head movements by the participant, can induce correlated noise between nearby voxels (Murphy et al., 2013; Power et al., 2012). Could a combination of these effects, rather than similarity in voxel tuning *per se*, explain the observed link between voxel tuning similarity and noise correlations? To test this, we repeated the analysis described above, with spatial proximity as an extra factor in the exponential decay functions that were fit to the data. We found that when this potential mediating factor was accounted for, the relationship between tuning similarity and shared noise remained generally significant (**Supplementary fig. 3**). Interestingly, this parallels results from macaque visual cortex (Bair et al., 2001; Smith and Kohn, 2008; Zohary et al., 1994), for which noise is preferentially shared between similarly tuned neurons.

Thus far, we have shown in simulations that forward models of fMRI activity that account for tuning-dependent noise correlations best characterize the probability distributions contained in data, and moreover, that such correlations are found in fMRI data from human visual cortex. Together, these findings suggest that by modeling tuning-dependent shared noise, it may be possible to decode probability distributions from cortical activity in human observers, thereby gaining insight into moment-to-moment fluctuations in the fidelity of cortical stimulus representations. Indeed, a decoder based on this principle can characterize the uncertainty in its stimulus estimates, while a naïve decoder fails to do so (**Fig. 3**; see also van Bergen et al. (2015)).

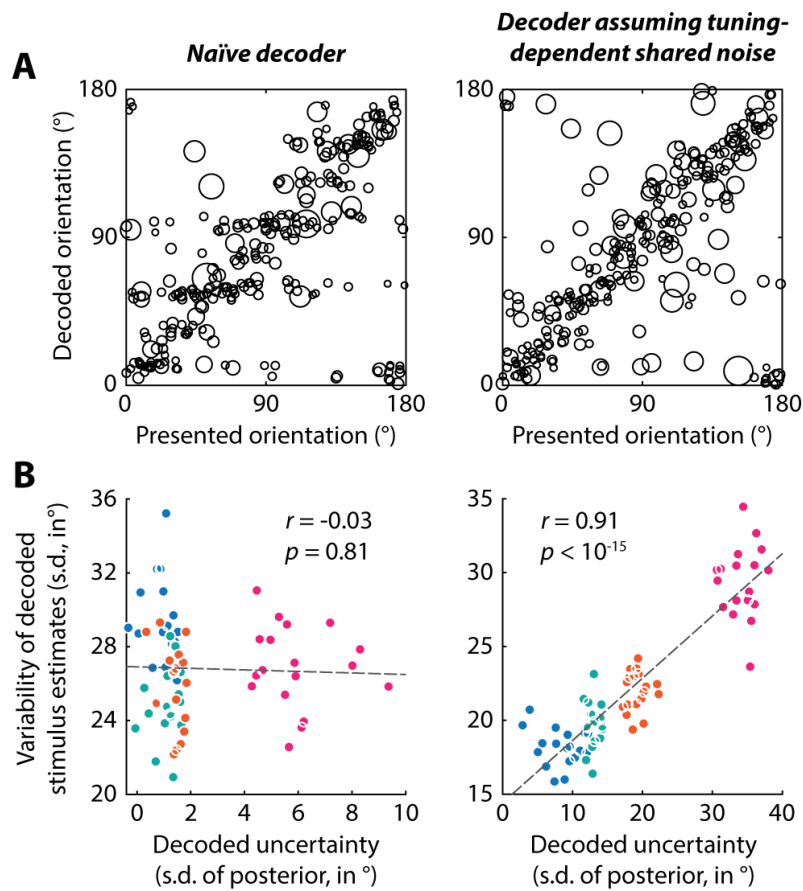

**Figure 3:** When tuning-dependent correlations are accounted for, uncertainty can reliably be estimated from fMRI data obtained from human visual cortex (data replotted from van Bergen et al. (2015). (**A**) Results from a representative subject, using a naïve decoder (left) and a decoder that models shared noise between similarly tuned voxels (right). Posterior distributions were estimated from the activity patterns evoked in visual cortex (areas V1-V3) by orientation grating stimuli. The circular mean of each distribution served as the decoded orientation estimate, while its width (circular standard deviation) indicated the decoded uncertainty in the activity pattern on that trial. Circles are centered on the decoded orientations and plotted against the presented stimulus orientations, while the size of each circle is proportional to the decoded uncertainty on that trial. For both decoding models, circles tend to lie near the diagonal (or in the far corners, due to the circularity of orientation space) indicating fairly accurate orientation estimates (across participants, the mean (circular) correlations between presented and decoded orientations were 0.57 and 0.69, for the naïve decoder and the one accounting for tuning-dependent shared noise, respectively). The decoder that incorporates shared noise captures the trial-by-trial uncertainty in the data reasonably well, as reflected by the greater dispersion of larger circles (greater uncertainty), compared to smaller circles, around the diagonal. The naïve decoder, on the other hand, fails to appropriately characterize the uncertainty in the data, as indicated by the fairly uniform distribution of circles around the diagonal, regardless of their size. (**B**) We quantified each decoder's ability to characterize uncertainty by relying on the relationship between uncertainty and variability (as the true uncertainty in real fMRI data is *a priori* unknown). More specifically, to the extent that the noise model appropriately captures the actual noise structure of the fMRI data, decoded uncertainty on a single trial should be linked to the variability in orientation estimates across trials. Trials were divided into 4 bins of increasing uncertainty, for each observer and for each of the two decoders. Colored dots show, for each participant, the mean decoded uncertainty, and mean variability of decoded stimulus estimates, across all trials in each bin, with bins indicated by different colors. The dashed line corresponds to the best linear fit of the data. The results obtained for the decoder incorporating tuning-dependent noise indicate a clear link between decoded uncertainty and across-trial variability in orientation estimates ($r = 0.91$, $p < 10^{-15}$). The naïve decoder, in contrast, failed to capture this relationship. This indicates that modeling shared noise is important for the decoding of uncertainty in fMRI activity patterns. See for further information, including a more detailed description of the decoding models and their analysis, van Bergen et al. (2015).

# 4 Discussion

We have shown that accounting for shared noise is important for forward decoding models of fMRI activity – without an explicit account of shared noise in the decoder, it is difficult to go beyond a mere prediction of the most likely stimulus, and assess the degree of uncertainty in the pattern of voxel activity. Specifically, our simulations demonstrate that probability distributions (that indicate uncertainty) become inaccurate when computed using 'naïve' decoders that ignore noise correlations in the data. We moreover find that a particular kind of correlated noise is most important to forward models of cortical activity, namely noise that is shared between voxels with similar tuning curves for the decoded stimulus feature. Including such tuning-dependent noise in the decoder's generative model enables the measurement of probability distributions that are very close to the true distributions, with widths that closely track the actual trial-by-trial uncertainty in the simulated activity patterns. In contrast, incorporating arbitrary noise correlations into the decoding model provided little improvement in decoding performance, even though these sources of noise contributed substantially and equally to the total amount of noise in the data. Interestingly, when analyzing fMRI data obtained from human visual cortex, we found that cortical noise correlations are also tuning-dependent. Moreover, and much in line with our simulations, a decoder that accounts for such tuning-dependent shared noise accurately predicted the trial-by-trial uncertainty in cortical activity patterns, while a naïve decoder failed to appropriately capture these moment-to-moment fluctuations in uncertainty.

Functional MRI only provides an indirect measure of neural activity (Lima et al., 2014; Logothetis et al., 2001), samples the aggregate signal from large neural populations rather than single cells, and is contaminated by many sources of noise, including scanner-related noise, physiological noise, and noise due to head or eye movements made by the participant (Greve et al., 2013; Henriksson et al., 2015; Kay et al., 2013; Power et al., 2012). An important next step would therefore be to establish the extent to which the decoded probability distributions indirectly reflect the uncertainty in neural population activity, as opposed to uncertainty due to noise in the fMRI measurements alone. In recent work, using similar tuning-dependent noise covariance models as discussed here, we took a first step in addressing this issue by correlating decoded distributions with subject behavior (van Bergen et al., 2015). Interestingly, our analyses revealed that decoded uncertainty was reliably linked with the accuracy of the observer's response, suggesting that the decoded probability distributions reflected, at least to some degree, the trial-by-trial uncertainty in underlying neural populations. The current study adds to this work by demonstrating the importance of different forms of correlated noise to this approach. This illustrates the necessary conditions for accurate probabilistic decoding from fMRI activity, and may facilitate the development of future probabilistic pattern analysis techniques in functional neuroimaging.

The majority of existing fMRI decoding algorithms do not explicitly account for noise correlations, but rather assume that variability is independent between voxels[1] (e.g. Kay, Naselaris, Prenger, & Gallant, 2008; Brouwer & Heeger, 2009; Serences, Saproo, Scolari, Ho, & Muftuler, 2009; Jehee, Ling, Swisher, van Bergen, & Tong, 2012; Ester, Anderson, Serences, & Awh, 2013). While such algorithms can provide reasonable estimates of the stimulus values presented to the observer, our results indicate that substantial additional information can be extracted from the data using a decoder that explicitly incorporates shared noise. Moreover, our results suggest that when constructing such a decoder, it is important to focus mostly on shared noise that resembles a tuning-based response. This observation may help to constrain the noise covariance structure in forward models of fMRI activity, by modeling specifically the most detrimental correlations.

---

[1] There are some notable exceptions to this general observation. For example, the optimal classification boundary in a support vector machine (SVM) depends on the contours of the voxel response distributions across trials, which are shaped differently when noise is correlated. These decoders, however, do not capture the full probabilistic relationship between stimuli and responses, and as such, are ill-equipped for measuring uncertainty.

Previous work has quantified the impact of noise correlations on the encoding and decoding of information across repeated presentations of the same stimuli (reviewed in Averbeck et al., 2006). For example, when evaluated across multiple presentations, tuning-dependent shared noise has been shown to decrease the mean precision with which a stimulus can be encoded in neural activity (Abbott and Dayan, 1999; Averbeck and Lee, 2006; Moreno-Bote et al., 2014; Smith and Kohn, 2008; Zohary et al., 1994). Similarly, the stimulus estimates of decoders that ignore noise correlations between like-tuned neurons are on average less accurate when evaluated across many stimulus presentations (Averbeck and Lee, 2006; Wu et al., 2001) – a finding we replicate for like-tuned fMRI voxels in **Fig. 1C**. By contrast, the focus of the current study is on the gamut of information contained in a *single* response pattern in cortex, characterized by a probability distribution over all possible interpretations of the cortical response. This ability to accurately extract the uncertainty of cortical information from a single neural population response is highly relevant to studies of cortical processing.

The relevance of uncertainty and probability distributions is perhaps best illustrated in the context of the ideal observer framework. Theoretically, it can be shown that decision-making agents achieve the best performance (i.e., the smallest errors or largest reward) if they use the uncertainty in their knowledge to calibrate their decisions. This allows the agent, for example, to discount in the decision-making process those sources of evidence that are unreliable. An increasing body of behavioral research suggests that the brain may operate like such an optimal agent (reviewed in e.g. Fiser et al., 2010; Knill and Pouget, 2004; Ma and Jazayeri, 2014; Pouget et al., 2013; Vilares and Körding, 2011). The ability to measure uncertainty with fMRI directly from human cortex enables investigations into the neural basis of such human decision-making under uncertainty (van Bergen et al., 2015). Being able to characterize uncertainty may also be advantageous when investigating internal neural variability *per se*. Although various external factors, such as a change in stimulus contrast, can be directly linked to uncertainty, some portion of neural response variability arises due to factors that are of internal origin (Faisal et al., 2008; Renart and Machens, 2014). The here presented decoding approach may open a window onto such stimulus-independent effects on information processing in cortex (van Bergen et al., 2015). Knowledge of uncertainty may also be beneficial to the design of brain-computer interfaces (BCIs). Imagine, for instance, a BCI that moves a robotic arm based on brain activity of a human operator. Due to various sources of noise, the reliability of the incoming brain signals will fluctuate over time. A BCI with knowledge of the uncertainty in these inputs could use this information to optimally integrate information across different dimensions of the incoming signal in order to arrive at the best possible interpretation of the intended motor command. Moreover, if overall levels of uncertainty turned out to be unacceptably high, the BCI could decide to delay executing the decoded motor command in order to avoid making erratic or unintended movements.

In conclusion, the here presented results show that including noise correlations into forward models of fMRI activity enables more accurate decoding of the gamut of information that is represented in a noisy pattern of data. Specifically, accounting for shared noise that mimics a stimulus-induced response provides a feasible method to characterize the moment-to-moment uncertainty in cortical activity, which we hope will open up new avenues for human neuroimaging research.

## Acknowledgements

This work was supported by ERC Starting Grant 677601 to J.J. We thank Wei Ji Ma, Christian Beckmann and Alberto Llera for helpful discussions, Kendrick Kay and John Serences for comments on an earlier draft of this paper, and Paul Gaalman for MRI support.

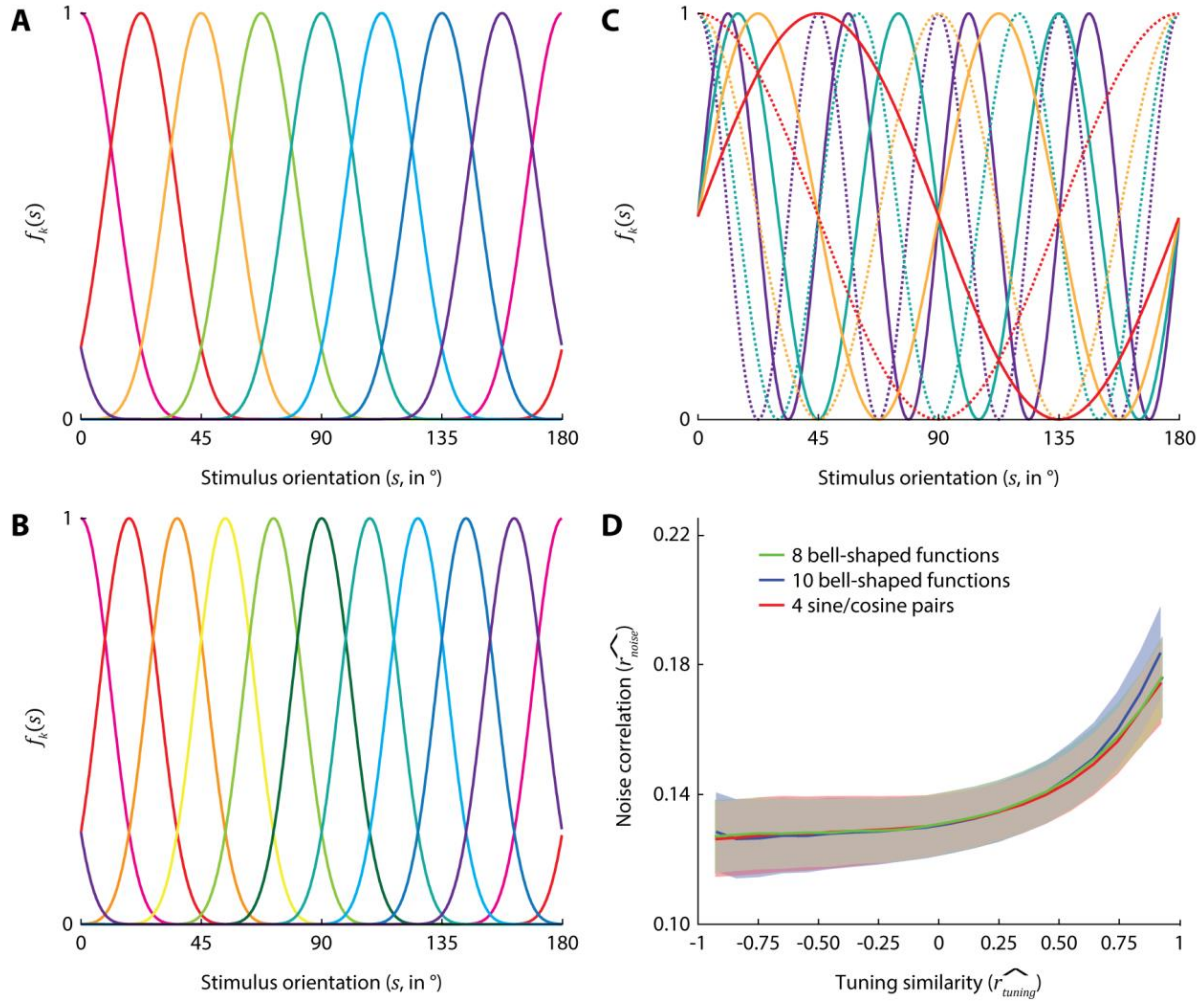

**Supplementary figure 1:** Estimated tuning similarity and noise correlations are robust to different choices of orientation basis functions used to fit voxel tuning curves.

(**A-C**) Three different basis sets used to fit voxel tuning curves. The value $f_k(s)$ of each ($k$-th) basis function is plotted against stimulus orientation ($s$). (**A**) The default basis set of eight bell-shaped functions (zero-rectified cosines, raised to the fifth power) used in the analyses presented in the main text. (**B**) A more expressive basis set of ten bell-shaped functions (zero-rectified cosines, raised to the seventh power). (**C**) An uncorrelated basis of four sines and four cosines, with periods of 180°, 90°, 45° and 25°. Sine/cosine pairs with the same period are shown in the same color, in solid/dotted lines, respectively. This basis spans the same space as the default set shown in **A**, but with orthogonal basis vectors. (**D**) Voxel tuning curves were fit as a linear combination of basis functions in each of the three basis sets show in **A-C**, and tuning similarity and noise correlations between pairs of voxels were calculated as in the main analysis (see **Methods section 2.8.1**). Lines and shaded regions indicate the mean +/- SEM noise correlation across subjects, in 20 equally spaced tuning similarity bins, for the three different basis sets (cf. **Fig. 2A**). Results are virtually identical regardless of which basis set is used to describe voxel tuning, indicating that slight correlations between adjoining functions in our chosen basis set only marginally affect the analyses, and that this basis provides sufficiently fine-grained coverage of orientation space.

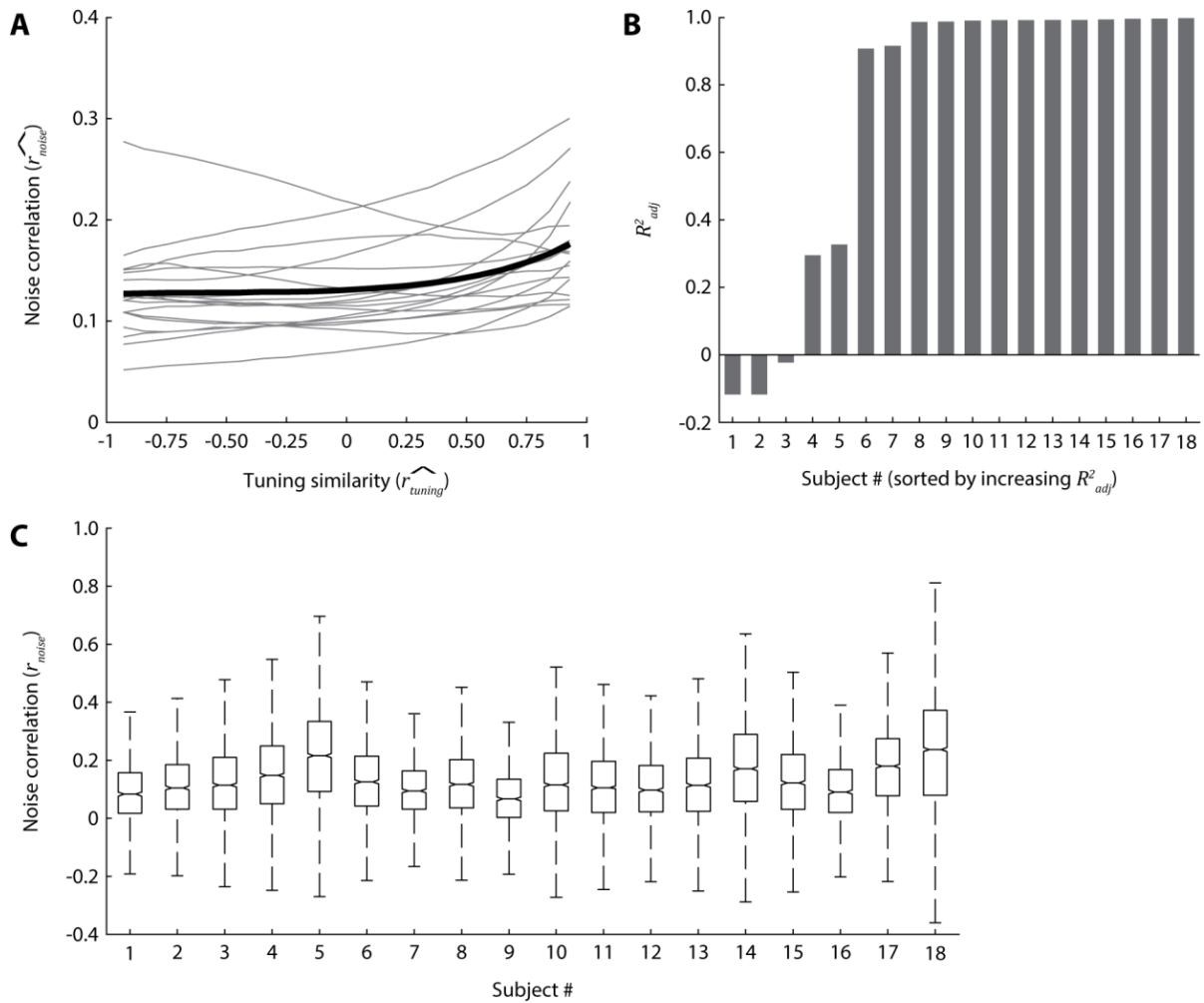

**Supplementary figure 2:** Inter-subject and inter-voxel variability in noise correlations.

(**A**) Average noise correlations ($\widehat{r_{noise}}$) plotted against average tuning similarity ($\widehat{r_{tuning}}$), in 20 equally space tuning similarity bins, for individual subjects (gray lines) and averaged across participants (black line; cf. **Fig. 2A**). (**B**) Goodness-of-fit per subject for the exponential decay function fits to the data in (**A**), quantified by means of the overfitting-adjusted coefficient of determination ($R^2_{adj}$). Subjects are sorted in order of increasing goodness-of-fit. For the majority of subjects (13/18), these data were extremely well fit by an exponential decay with decreasing tuning similarity, with overfitting-adjusted $R^2$-values in excess of 0.9. Fits were moderately good for another two subjects ($R^2_{adj}$ around 0.3), and below 0 for the remaining three ($R^2$-values below 0 arise after adjustment for overfitting). (**C**) Box plots showing the distribution of noise correlations across voxel pairs, per subject. Boxes extend from the first to the third quartiles, with notches indicating the medians. Whiskers span the range between the most extreme data points that were not selected as outliers. Outliers were defined as values deviating from the mean by more than 1.5 times the inter-quartile range. For figure clarity, outliers are not shown, due to the large number of data (+/- 2 million voxel pairs per subject). Note that subject order here is not the same as in (**B**).

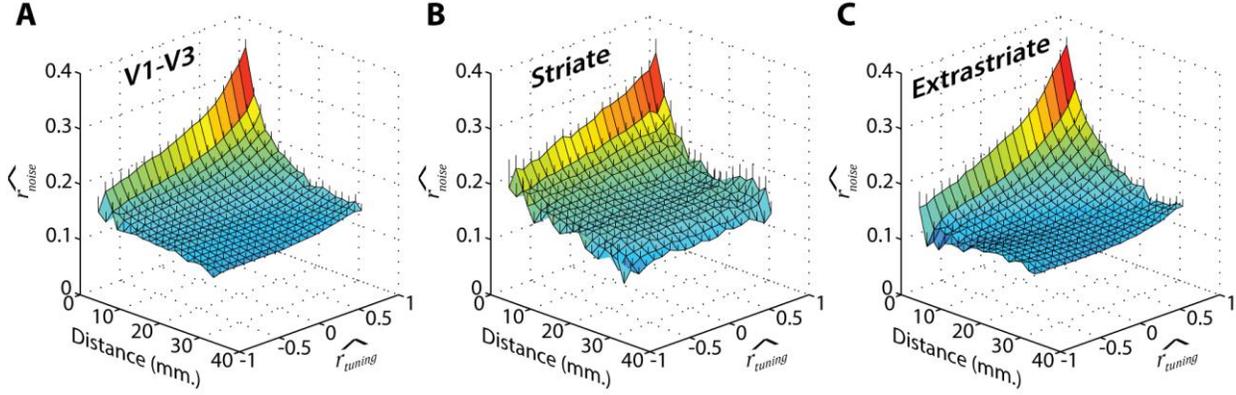

**Supplementary figure 3:** Shared noise depends on both orientation tuning and inter-voxel distance.

Given that neighboring voxels in cortex tend to have similar orientation tuning preferences (Freeman et al., 2011; Mannion et al., 2010; Sasaki et al., 2006; Swisher et al., 2010), do noise correlations arise because of noise that is shared between neighboring voxels (e.g., due to the BOLD point spread function (Parkes et al., 2005) or other causes of spatially correlated noise; see e.g. Arcaro et al. (2015), Henriksson et al. (2015), Murphy et al. (2013), and Power et al., (2012)), rather than noise that depends on tuning *per se*? To address this question, voxels pairs were sorted into 20 x 20 bins of similar inter-voxel distance and tuning similarity ($\widehat{r_{tuning}}$). Within each bin, the mean noise correlation, tuning similarity and distance across all pairs of voxels was calculated. This produced, for each observer, a three-dimensional noise correlation surface. Panels (**A-C**) show the average surface across subjects for V1-V3 combined, striate (V1) and extrastriate (V2-V3) cortex, respectively. Group-average surfaces were calculated by taking the Fisher-transform of the surface for each subject, averaging across subjects, and then transforming these averages back to the correlation scale. Lines protruding vertically out of the mesh surfaces indicate + 1 SEM.

To quantify the degree to which noise correlations can be explained by spatial distance and/or tuning similarity, noise correlation surfaces obtained for each participant were fitted with each of four models. The models described noise correlations as a function of decreasing tuning similarity (model 1), increasing distance (model 2) or both (models 3 and 4). More specifically, model 1 describes an exponential decay in noise correlations with decreasing tuning similarity (cf. equation (17)):

$$h(m,n) = \alpha \exp\left(-\beta\left(1 - \langle\widehat{r_{tuning}}\rangle_{m,n}\right)\right) + \gamma \tag{S1}$$

where $h(m,n)$ is the predicted noise correlation for voxel pairs in the $(m,n)$-th bin ($m$ indexes bins of different tuning similarity and $n$ enumerates bins of different inter-voxel distance), $\langle\widehat{r_{tuning}}\rangle_{m,n}$ is the mean (estimated) tuning similarity between voxel pairs in the $(m,n)$-th bin, $\beta$ and $\alpha$ control the rate and starting value of the decay, and $\gamma$ models a constant baseline correlation among all voxels.

Model 2 describes noise correlations as decaying exponentially with increasing inter-voxel distance:

$$h(m,n) = \kappa \exp(-\lambda\langle d\rangle_{m,n}) + \gamma \tag{S2}$$

where $\langle d\rangle_{m,n}$ is the average inter-voxel distance in bin $(m,n)$, and $\lambda$ and $\kappa$ determine the rate and starting value of the decay.

Model 3 describes noise correlations as decaying with both increasing distance and decreasing tuning similarity:

$$h(m,n) = \alpha \exp\left(-\beta\left(1 - \langle\widehat{r_{tuning}}\rangle_{m,n}\right)\right) + \kappa \exp\left(-\lambda\langle d\rangle_{m,n}\right) + \gamma \quad (S3)$$

Model 4 describes the decline in noise correlations as an interaction between an increase in distance and a decrease in tuning similarity:

$$h(m,n) = \alpha \exp\left(-\beta\left(1 - \langle\widehat{r_{tuning}}\rangle_{m,n}\right)\right) \exp\left(-\lambda\langle d\rangle_{m,n}\right) + \gamma \quad (S4)$$

Each of these exponential decay functions were fit to the data by minimizing the squared error between the Fisher-transformed predictions and the Fisher-transformed bin-average noise correlations, and decay amplitudes ($\alpha$ and $\kappa$) and rates ($\beta$ and $\lambda$) were constrained to be non-negative. Goodness-of-fit was assessed by calculating the overfitting-adjusted coefficient of determination ($R^2_{adj}$) for each model and subject, and then a one-tailed Wilcoxon-signed rank test was used to determine whether the average variance explained across subjects ($\langle R^2_{adj}\rangle$) was reliably greater than zero, for each model. Between models, goodness-of-fit was compared using a two-tailed Wilcoxon-signed rank test.

Although tuning similarity by itself explained significant variance in the noise correlation surfaces (model 1: V1-V3 combined: $\langle R^2_{adj}\rangle$ = 0.22, $p < 10^{-4}$; striate cortex: $\langle R^2_{adj}\rangle$ = 0.19, $p$ = 0.041; extrastriate cortex: $\langle R^2_{adj}\rangle$ = 0.27, $p < 10^{-4}$), these data were best described by an interaction between tuning similarity and distance in extrastriate cortex and V1-V3 combined (model 4: extrastriate cortex: $\langle R^2_{adj}\rangle$ = 0.83, $p < 10^{-5}$; V1-V3 combined: $\langle R^2_{adj}\rangle$ = 0.84, $p < 10^{-5}$; comparisons between model 4 and all other models, in both ROIs: all $p < 0.01$). In striate cortex, a combination of tuning and distance better described the data than tuning alone (model 3: $\langle R^2_{adj}\rangle$ = 0.65, $p < 10^{-5}$; model 4: $\langle R^2_{adj}\rangle$ = 0.64, $p < 10^{-5}$; comparisons of models vs. 3 & 4 vs. model 1, both $p < 10^{-5}$), and this combination was marginally better when compared with a model in which noise correlations decayed with increasing distance alone (model 2 vs. 3, $p$ = 0.074; model 2 vs. 4, $p$ = 0.054). Together, these results indicate that shared fMRI response variability in areas V1-V3 depends on both the distance between voxels and their orientation tuning properties.

Please note that, from a *decoding* perspective, precisely how tuning-dependent correlations arise is of lesser importance: as long as tuning-dependent correlations are present in the data, decoding performance will improve when the generative model takes them into account (as exemplified by the simulations in the main text).